\documentclass[prl,twocolumn,aps,showpacs]{revtex4}

\usepackage{here}
\usepackage{graphicx}
\graphicspath{{pict/}{}}

\newcommand{\be}{\begin{equation}}
\newcommand{\ee}{\end{equation}}

\newcommand\pictc[5]{\begin{figure}[t,b]
                       \centerline{\vspace{-0mm}
          \includegraphics[width=#1\columnwidth,height=0.7\textheight,keepaspectratio]{#3}}
                       \protect\caption{\protect\label{fig:#4} #5}\vspace{-0mm}
                    \end{figure}            }

\newcommand\pict[4][1]{\pictc{#1}{}{#2}{#3}{#4}}
\newcommand\rpict[1]{\ref{fig:#1}}

\newcounter{Fig}

\begin{document}
\begin{sloppy}

\title{Generation of the second-harmonic Bessel beams via nonlinear Bragg diffraction}

\author{Solomon~M. Saltiel$^{1,2}$, Dragomir~N. Neshev$^1$, Robert Fischer$^1$,
Wieslaw~Krolikowski$^1$, Ady Arie$^3$, and Yuri~S.~Kivshar$^1$}

\affiliation{$^1$Nonlinear Physics Center and Laser Physics Center,
Center for Ultra-high bandwidth Devices for Optical Systems (CUDOS),
Research School of Physical Sciences and Engineering, Australian
National University, Canberra ACT 0200, Australia}

\affiliation{$^2$Department of Quantum Electronics, Faculty of
Physics, Sofia University, Bulgaria}

\affiliation{$^3$School of Electrical Engineering, Faculty of
Engineering, Tel-Aviv University, Ramat Aviv, Tel-Aviv, Israel}

\begin{abstract}
We generate conical second-harmonic radiation by transverse
excitation of a two-dimensional annular periodically-poled nonlinear
photonic structure with a fundamental Gaussian beam. We show that
these conical waves are the far-field images of the Bessel beams
generated in a crystal by parametric frequency conversion assisted by  nonlinear Bragg diffraction.
\end{abstract}


\maketitle

The second-harmonic generation, i.e. conversion of two photons of
the fundamental wave into a single photon at twice the frequency,
belongs to one of the most intensively studied nonlinear effects
\cite{Franken:RMP:1963}. Such a process takes place in quadratic
optical media, i.e. nonlinear media without the center of inversion.
It is well established that the efficiency of this as well as other
similar parametric processes depends critically on the so-called
phase-matching condition which in birefringent crystals is commonly
achieved with the temperature or angle tuning. In media with weak
birefringence, an efficient frequency conversion is achieved by the
quasi-phase matching (QPM)~\cite{QPM}. The QPM technique involves periodic
modulation of the second-order nonlinearity of the material. In the
ferroelectric crystals such as LiNbO$_3$ or LiTiO$_3$,  this can be
easily realized by spatially periodic poling~\cite{poling}. In this
way one can create one- or two-dimensional nonlinear structures, the
so-called  $\chi^{(2)}$ photonic crystals
\cite{Berger:PRL:98,Xu:PRL:2004}. Depending on the symmetry and form
of the poling pattern the efficient second-harmonic generation (SHG)
can be realized for waves propagating in particular spatial
directions~\cite{Kurz:JSTQE:02}. Typical geometries involve, in one
spatial dimension, single or multi-period rectangular pattern, and,
in two spatial dimensions,  nonlinear photonic structures of
hexagonal or square symmetries.

In this Letter, we report on the generation of conical
second-harmonic waves by excitation of a two-dimensional annular
$\chi^{(2)}$ nonlinear photonic fabricated in Stoichiometric Lithium
Tantalte (SLT) crystal with a single Gaussian beam. We show that the
observed rings of the second-harmonic radiation actually represent
the far field of the Bessel beams generated in the crystal via the
multi-order nonlinear Bragg diffraction.

We consider a two-dimensional nonlinear photonic structure with
circular periodic array of ferroelectric domains with a constant
linear refractive index. In general, such two-dimensional QPM
photonic structures have been explored exclusively in the {\em
longitudinal}
geometry~\cite{Berger:PRL:98,Broderick:PRL:2000,Xu:PRL:2004,Ady:OE:2006}
of the second-harmonic generation when the fundamental beam
propagates perpendicular to the domains boundaries of periodically varying
second-order nonlinearity. Here, we explore a novel {\em transverse}
geometry when the fundamental light beam propagates along the axis
of the structure. The front facet of the annular poled
Stoichiometric Lithium Tantalate sample used in our experiments with
clearly visible domain boundaries is shown in Fig.~1(a) [for
technical details, see \cite{Ady:OE:2006}].  It is a Z-cut,
$L=0.49$\,mm thick slab with the QPM period of $7.5\,\mu $m and duty
factor varying  inside the sample from 0.7 at Z+ surface to 0.8 at
Z-surface. Both, Z+ and Z- surfaces are polished. Small 20\,nm deep
grooves that remain  after polishing cause weak (less than 3{\%})
diffraction of the fundamental beam which we send perpendicular to
the domain structure, along its axis. As a light source we use a
Ti:Sapphire oscillator and regenerative amplifier delivering beam of
140\,fs pulses, with a repetition rate of 250\,kHz and average
output power of 740\,mW. Time bandwidth product of the laser pulses
is 0.49.

The input beam  is loosely focused such that  the beam waist at the
input facet of the sample  is 147\,$\mu$m (FWHM) which, for a given
input power, corresponds to the intensity about 100\,GW/cm$^2$. The
beam covers roughly 10 domain rings. We observe that the transverse
illumination of the radial structure by the fundamental beam lead to
multiple conical emission of the second-harmonic (SH) waves [see
Figs.~1(b) and ~2(a)]. These waves form a set of rings in the far
field, and their propagation angles satisfy the Bragg relation, i.e
they are determined by the ratio of the wavelength  of the second
harmonic to the period of the modulation of the $\chi^{(2)}$
nonlinearity.

To explain the appearance of the second-harmonic rings, we consider
the phase-matching conditions~\cite{SaltielOE} for the corresponding
parametric process shown in Fig.~\rpict{scheme}(c). The general
vectorial phase-matching condition can be split into two scalar
ones: the transverse condition, $k_2 \sin \alpha_m = G_m = mG_1$,
and the longitudinal condition, $k_2 \cos \alpha_m - 2k_1 = \Delta
k_m$. With a correct choice of the fundamental wavelength or
temperature, a given diffraction order will be longitudinally
phase-matched ($\Delta k_m =0$). As an example, in
Fig.~\rpict{scheme}(c) the third diffraction order is phase-matched
($\Delta k_3 =0$), and for this interaction  both the phase-matching
conditions are satisfied simultaneously. Inside the crystal, the
propagation angles of all transversely phase-matched waves can be
found from the relation $\sin \alpha_m = m\lambda_2 / n_2 \Lambda$,
which is analogous to the Bragg condition governing the diffraction
of waves on periodic modulation of the index of refraction.  Here
the refractive index is constant but the nonlinear properties of the
medium experience spatially periodic variations. This effect is
known as {\em nonlinear Bragg diffraction}, and it was first
discussed by I. Freund in 1968~\cite{Freund:PRL:1968}.

The condition of the total internal reflection determines the
maximum number of the second-harmonic rings,
$M_1=\Lambda/\lambda_2$. The SHG process in the transverse direction
($\beta_m=90^\circ$) determines the total maximum diffraction
orders, $M_2=n_2\Lambda/\lambda_2$. For this transverse SHG process
the longitudinal mismatch is $\Delta k_m=2k_1$, and its compensation
is possible only when two counter-propagating fundamental waves are
employed~\cite{FischerAPL2007}.

\pict{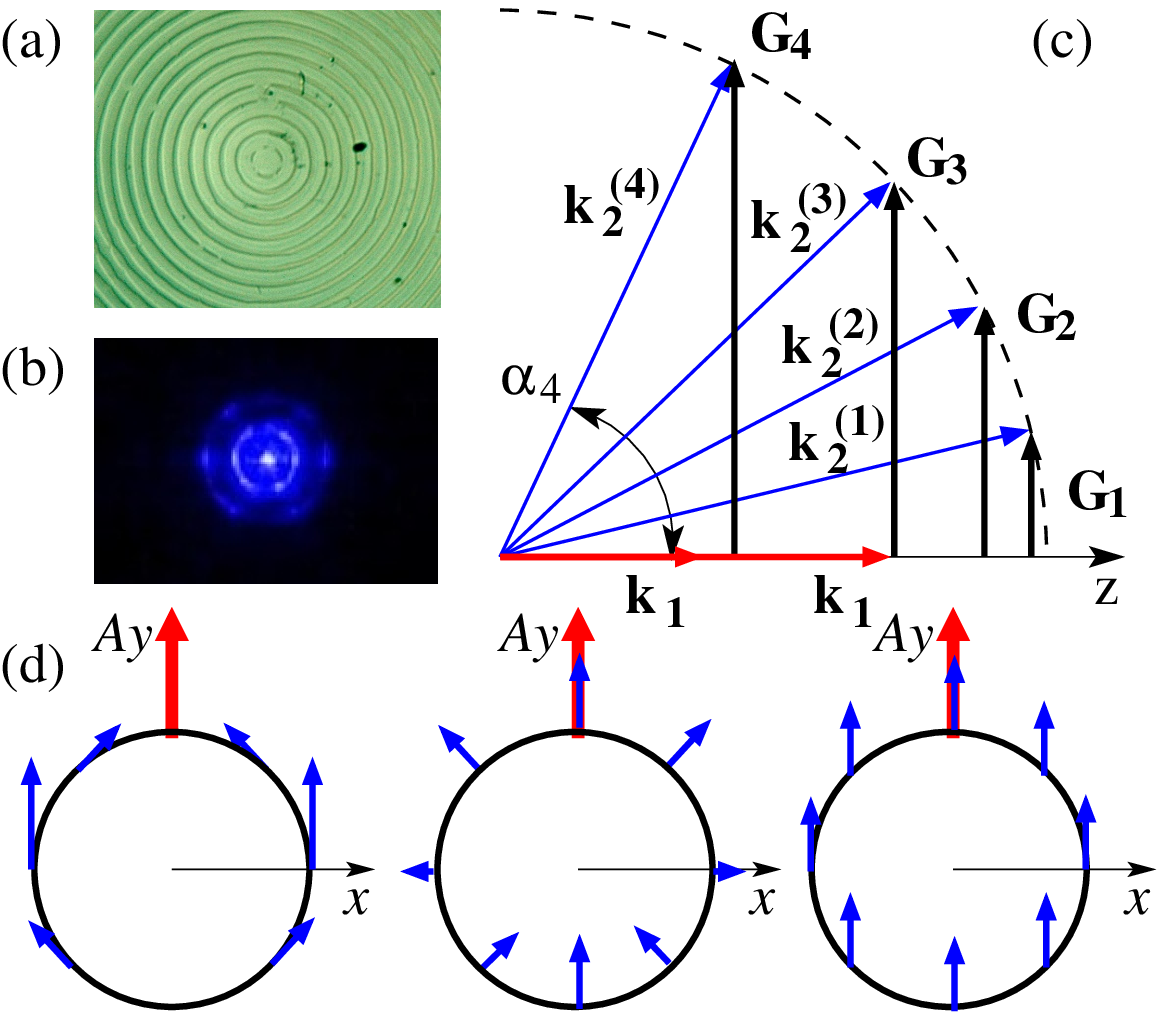}{scheme}{(color online) (a) Front facet of the SLT
sample. (b) Far-field image of first two diffraction SH rings. (c)
Phase matching diagram of the SHG process in the transverse QPM
grating. (d) Polarization structure of the ordinary (left),
extraordinary (middle), and small-angle (right) rings. The
fundamental wave (red) has $Y$ polarization.}

For the annular periodically poled structure, the outer rings
possess interesting polarization properties determined by the tensor
of the second-order nonlinearity of the Lithium Tantalate. This
crystal belongs to the 3m symmetry point group. For the fundamental
beam propagating along the $Z$ axis, there exist following nonzero
components: $d_{\rm zxx} = d_{\rm zyy}, d_{\rm yxx} = d_{\rm xyx}= -
d_{\rm yyy}$. Note that in periodically poled crystals with 3m
symmetry the values $d_{\rm xxy}, d_{\rm yxx}, d_{\rm yyy}$ also
change signs periodically~\cite{Ganany}. Two types of the SHG
processes are possible: Type I (e-oo, two ordinary waves generate an
extraordinary second-harmonic wave) and Type 0 (o-oo, two ordinary
waves generate an ordinary second-harmonic wave). Effective
nonlinearities for these two processes are~\cite{Zernike}
\begin{equation}
\label{eq1} d_{\rm eff}^{(o)}=d_{\rm yyy}\cos(\varphi + 2 \gamma),
\end{equation}
\begin{equation}
\label{eq2} d_{\rm eff}^{(e)}=d_{\rm yyy}\cos \alpha \sin(\varphi +
2 \gamma) + d_{\rm zyy}\sin \alpha,
\end{equation}
where $\varphi$ is the azimuthal angle measured counterclockwise  from the $X$ axis,
$\alpha$ is the diffraction angle inside the crystal, and $\gamma$ is
the angle that defines the polarization of the input beam
 ($\gamma =0$ for polarization  along $X$ axis).
 The azimuthal intensity distribution of the
``ordinary'' and ``extraordinary'' second-harmonic rings is given by
$I_{2\omega}^{(o,e)}\propto \left [ d_{\rm eff}^{(o,e)} \right ]^2 g_m^2
I_1^2 L_{\rm coh}^2$. Hence, the total intensity is
\begin{equation}
\label{total_int} I_{2\omega,m} \propto \left\{ \left [ d_{\rm
eff}^{(o)} \right ]^2 + \left [ d_{\rm eff}^{(e)} \right ]^2
\right\} g_m^2 I_1^2 L_{coh,m}^2,
\end{equation}
where $I_1$ denotes the intensity of the fundamental beam,
$g_m=[2/(\pi m)]sin(\pi m D)$, $L_{coh,m}=\pi/\Delta k_m$ and $D$ -
represents the duty factor. The light intensity in individual rings
depends also on the phase mismatch $\Delta k_m$. For ring(s) for
which both the phase-matching conditions are fulfilled the coherence
length $L_{\rm coh}$ has to be replaced by the actual crystal length
$L$. In Fig.~2(b) we illustrate the dependence of the relative
intensity (in log scale) in each diffraction order on $m$.

\pict{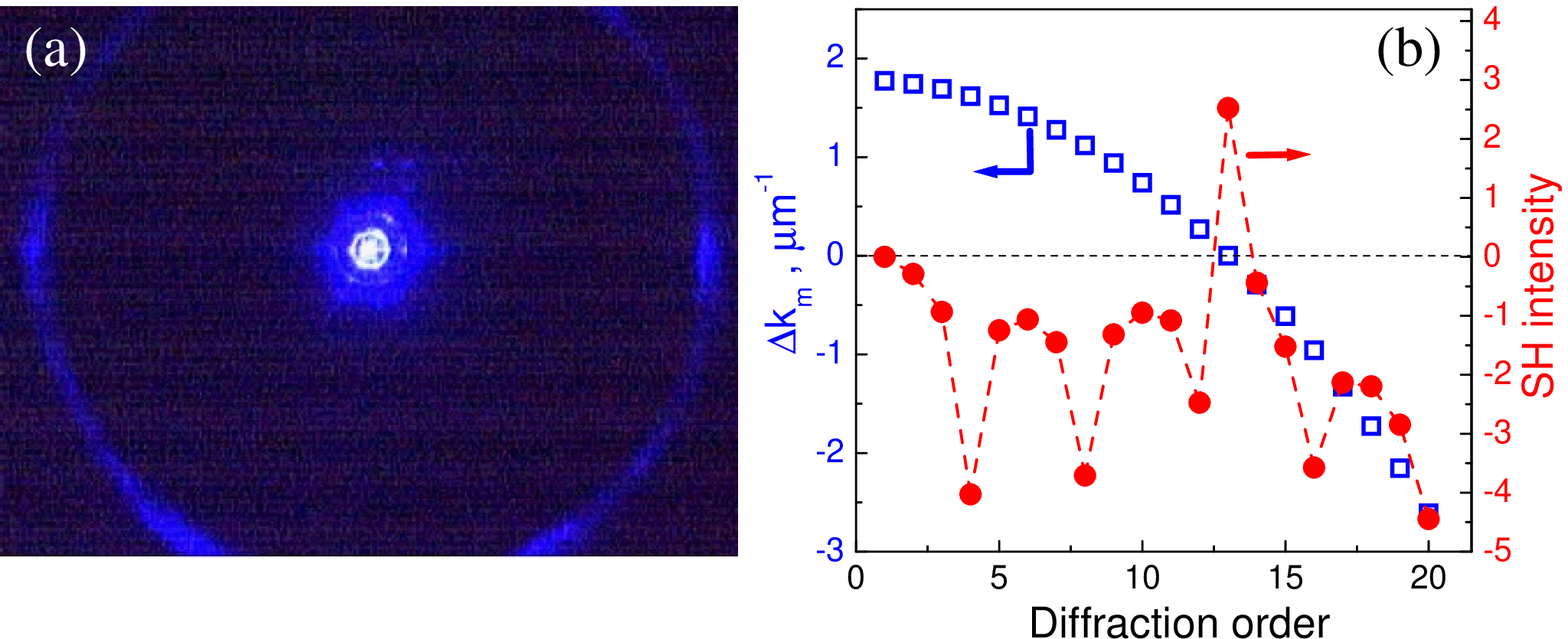}{rings}{(color online) Far field image of
experimentally observed second-harmonic rings. The axis $X$ is
horizontal. (a) Fundamental beam is exactly in the center of the
annular structure: first several orders and 13-th order SH ring are
seen. (b) Dependencies of the wave vector mismatch and the predicted
SH efficiency normalized to the first-order efficiency (in log
scale) as a function of the diffraction order.}

For small angles $\alpha$ (when $\tan\alpha <<1$), the difference in
the refractive indices of the ordinary and extraordinary  waves can
be neglected, the ``ordinary'' and ``extraordinary'' SH rings
perfectly overlap, and as a result the intensity  of conically
emitted light does not depend on $\gamma$ and $\varphi$. In
addition, its polarization becomes linear [Fig. 1(d, right)]. For
$\gamma =0$  and for $\gamma =\pi/2$ polarization of the   SH wave
 coincide with the $Y$-direction (with relevant nonlinear
components $d_{yxx}$ and $d_{yyy}$, respectively), while for $\gamma
=\pi/4$  polarization of the   SH wave coincide with the $X$-direction
(with relevant nonlinear  component $d_{xyx}$).

For relatively large $\alpha$, the ordinary and extraordinary waves
propagate with different phase velocities [$n_o > n_e(\alpha)$] and
at slightly different conical  angles. In this case, the output
polarization depends strongly on the actual spatial overlap of the
two rings since they are orthogonally polarized. In
Fig.~\rpict{scheme}(d) we show the structure of the predicted
polarization on the ring for $d_{yyy}/d_{zyy} \approx
-1.7$~\cite{Landolt} and $\alpha=20^\circ$. For relatively long
crystals every diffraction order should appear as a doublet
consisting of two orthogonally-polarized rings.

A full analysis of the SHG process in  the QPM structure with the
azimuthal symmetry shows that, in fact, the SH field inside the
crystal has a form of a nondiffracting Bessel beam~\cite{bessel}.
Detailed form depends on the crystal symmetry. In the considered
earlier case of Strontium Barium Niobate, the SH wave is just the
first-order Bessel function with a perfect radial
polarization~\cite{SaltielOE}. On the other hand, in case of
LiTaO$_3$ or LiNbO$_3$, the beam structure is more complicated due
to the contributions of several components of the $\chi^{(2)}$
tensor. However, it still can be represented in terms of a
superposition of the low-order Bessel functions with the angularly
modulated intensity. The exact formulas are quite cumbersome, but
they can be simplified in case of small emission angles ($\tan\alpha
<<1$),
 \be\label{xfield}
  E(\rho,\theta_0,z)\propto 2\pi A^2d_{\rm yyy}J_0(\xi)\left[
   \hat{u}_x\sin2\gamma +\hat{u}_y\cos2\gamma \right],
\ee
where $\xi=k_2\rho\sin\alpha$ and $\rho=(x^2+y^2)^{1/2}$ is the
transverse radial coordinate. $\hat{u}_x$ and $\hat{u}_y$ are unit
vectors along $X$ and $Y$. It is clear that the beam represented by
Eq.~(\ref{xfield}) is linearly polarized, with its intensity being
independent of the azimuthal coordinate $I \propto 4\pi^2d^2_{\rm
yyy}A^4J^2_0(\xi)$.  This results are  in a full  agreement with
experimental observations [Fig.~\rpict{shift_intdist}(d)].

\pict{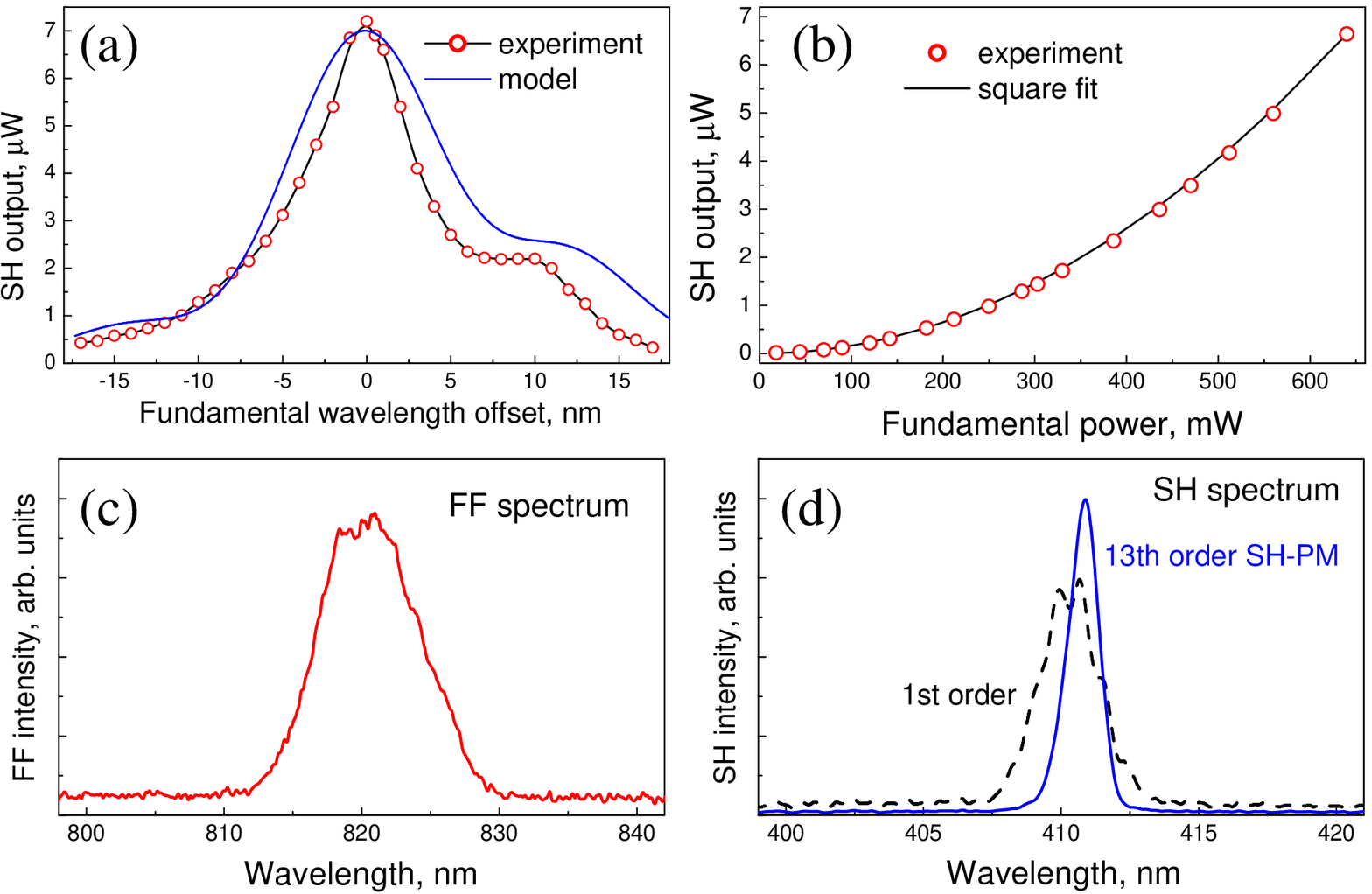}{quadr_PM_spectra}{(color online)  (a)
Experimentally  measured  (circles) and theoretically predicted
(solid line) phase matching curves. (normalized to the maximum
signal of the experiment). (b) Quadratic dependence of the 13-th
order SH ring on input power. (c) Spectrum of the fundamental pulse.
(d) Spectra of the SH rings  for non phase matched (first-order) and
phase matched (13-th order) SH ring.}

The observed SH rings for the fundamental wavelength of  822\,nm are
shown in Fig.~\rpict{rings}(a). The measured conical angles agree
well with those determined  from the nonlinear Bragg diffraction
formula. By re-scaling the size of the lower-order diffraction
pattern we  found that the phase matched SH emission which is
represented in the far-zone by the  ring with the largest diameter
and conical angle of $45.6^{\circ}$ corresponds to the high 13-th
diffraction order. In Fig.~\rpict{rings}(b) we depict the calculated
(using the published refractive index data \cite{Nakamura})
longitudinal phase mismatch and relative power of the SH waves for
$\lambda=836$\,nm. The highest peak  corresponds to the
phase-matched ($\Delta k_{13}=0$) 13-th order SH ring.

The phase-matching curve is shown in
Fig.~\rpict{quadr_PM_spectra}(a) together with the theoretical
prediction based on the model from Ref.~\cite{SaltielOE}. We observe
good qualitative agreement between the experimental curve and the
theoretical  prediction. In Fig.~\rpict{quadr_PM_spectra}(b) we
verify a quadratic dependence of the SH power of the phase matched
13-th SH ring as a function of the input power. From this figure
(obtained at 640\,mW fundamental average power taking into account
all losses), we obtain the internal efficiency for the phase-matched
emission of $\eta_{13}=0.0056${\%} / W. Such low efficiency is
attributed to the fact that we use  high order QPM interaction.
Additionally, the variations of the duty factor inside the crystal
leads to an effective length 2-3 times less than the real sample
thickness. The measured efficiency of the SH power generated in the
first-order ring is 220 times less than $\eta_{13}$. The spectra are
shown in Fig.~3(c,d). The fundamental spectral width is 8\,nm. The
spectra of the SH ring in the case of phase-matching of the 13-th
order is 1.3\,nm; i.e. almost 2 times less than  expected for a
stationary process. This indicates  that the SHG process takes place
in the group-velocity mismatch conditions. Indeed, the calculated
group-velocity-mismatch length is 0.092\,mm for 140\,fs fundamental
pulse width, that is a half of the expected effective length of the
sample. Subsequently, the output spectra (pulse length) is expected
to be reduced (stretched) by a factor of 2. In contrast, the spectra
of non-phase matched rings [see Fig. 3(d)] do not show any reduction
and are close to those predicted for the stationary process,
$\Delta\omega_{\rm SH} = \sqrt{2}\Delta\omega_{\rm FF}$.

\pict{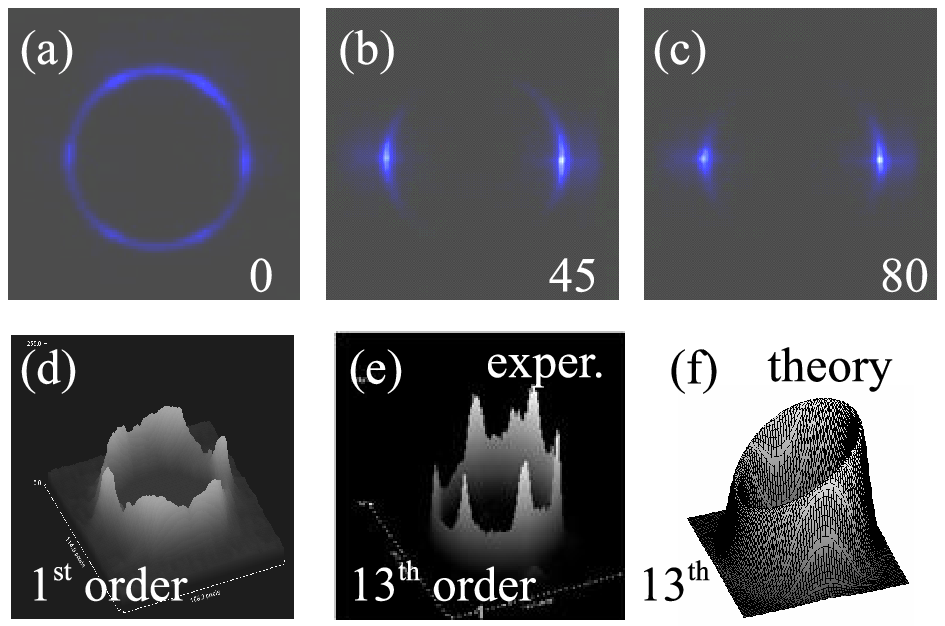}{shift_intdist}{(color online) (a-c)
Modification of the diffraction pattern as a function of the
horizontal misalignment of the incident beam. Numbers indicate the shift in $\mu$m. The
crystallographic  $X$ axis is horizontal. (d,e) Experimentally recorded intensity
distribution of $m=1$ and $m=13$ SH rings. (f) Theoretical
prediction for the azimuthal intensity distribution of the $m=13$ SH
ring evaluated from  Eq.~(3).}

\pict{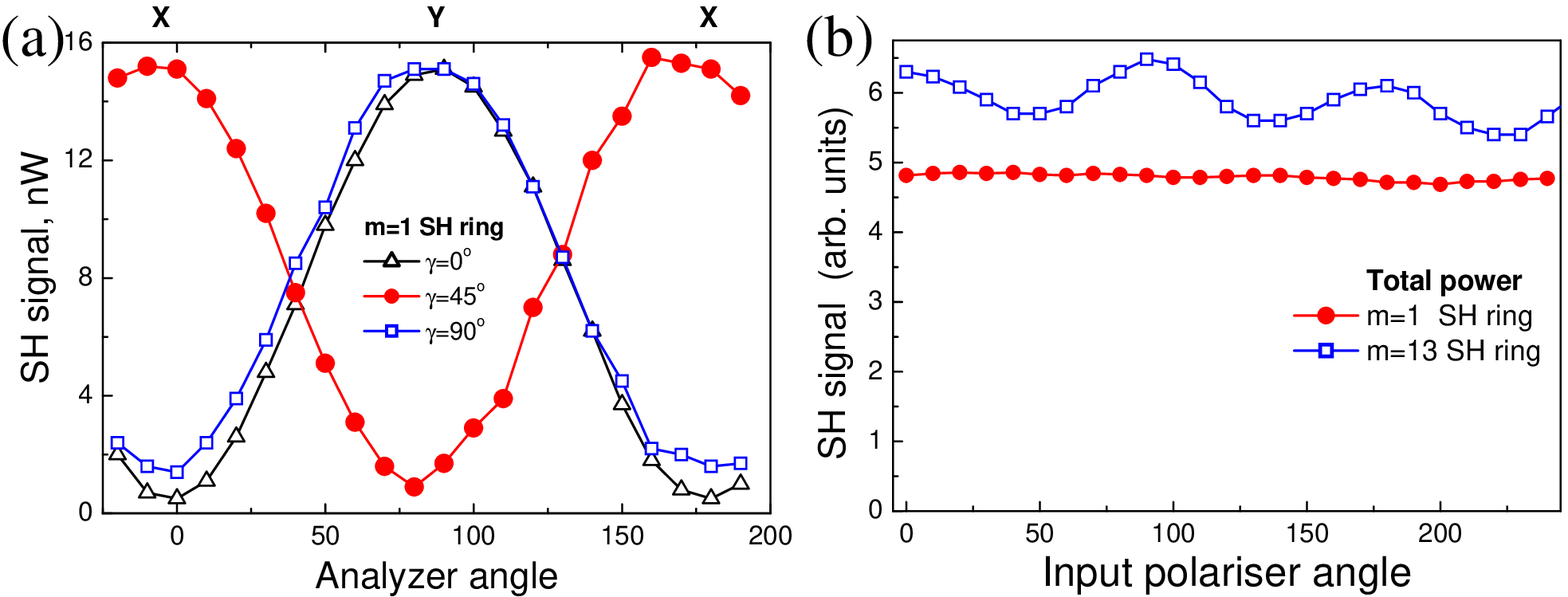}{pol_dep}{(color online) Polarization properties
of the SH rings: (a) Power of the 1-st order SH ring vs. analyzer
angle, (b) Total power of the 1-st and 13-th order SH rings vs. the
input polarization angle.}

The rings are sensitive to the axial alignment of the fundamental
beam. With the off-center beam, the rings transform into a two-peak
structure, the efficiency increases, and the number of the observed
diffracted orders increases too. For a horizontal shift of the
sample, the diffraction spots appear horizontally. For a vertical
shift of the sample, the diffraction spots appear vertically. This
is explained by the fact that the structure resembles a
one-dimensional grating formed from the sectors situated close to
the horizontal (vertical) diameter of the rings. The modifications
of the diffraction pattern for the phase-matched 13-th order ring in
the case of off-center excitement are shown in
Figs.~\rpict{shift_intdist}(a-c). Figures~\rpict{shift_intdist}(d,e)
show the experimental intensity distribution of the 1-st and 13-th
order rings. The theoretical prediction for the 13-th order ring is
given by Eq.~(\ref{total_int}), and it is shown in
Fig.~\rpict{shift_intdist}(f).

As follows from Fig.~1(b) and Figs.~\rpict{shift_intdist}(a,d,e), the
second-harmonic rings exhibit an additional azimuthal modulation
with 6 well defined peaks. These peaks are insensitive to the input
polarization and intensity, and they appear even in the regime of
linear diffraction of the fundamental beam on the 20 nm deep surface
relief - replica of poling pattern. The peaks are artifacts of  the
discrete character of domain boundaries resulting from the well
known tendency of ferroelectric domains in the SLT to retain their
hexagonal shape~\cite{Lobov,Ady:OE:2006} .

In Fig.~5 we show our results of the polarization measurements of
the second-harmonic rings. For the linearly polarized input
fundamental beam the generated  wave is also linearly polarized in
the case of  the first-order ring. This result agrees with Eqs.~(1)
and (2) in the limit $\alpha \rightarrow 0$. Polarization state
becomes more complex for higher-order nonlinear diffraction as the
output signal contains  both ordinary and extraordinary
contributions.

In conclusion, we have reported on the first observation of the
second-harmonic Bessel beams  generated by two-dimensional annular
nonlinear photonic structures. We have explained the effects
observed in experiment by employing the concept of nonlinear Bragg
diffraction combined with the longitudinal and transverse phase
matching conditions. We have studied the polarization properties of
the conical second-harmonic radiation and describe how the beam
polarization depends on the crystal symmetry and conical angle.

This work was supported by the Australian Research Council and the
Israeli Science Foundation (grant no. 960/05). We thank D. Kasimov,
A. Bruner, P. Shaier, and D. Eger for their assistance in the
fabrication of the SLT samples. S. Saltiel thanks Nonlinear Physics
Center for hospitality and support.


\end{sloppy}

\begin{thebibliography}{22}
\bibitem{Franken:RMP:1963}P. A. Franken and J.F. Ward, Rev. Mod. Phys. {\bf 35},  23 (1963).

\bibitem{QPM}M.M. Fejer, G.A. Magel, D.H. Jundt, and R.L. Byer,
IEEE J. Quant. Electron. {\bf QE-28}, 2631-2654 (1992).

\bibitem{poling}M. Houe and P. D. Townsend,
J. Phys. D: Appl. Phys. {\bf 28}, 1747 (1995);

\bibitem{Berger:PRL:98} V. Berger, Phys. Rev. Lett. {\bf 81}, 4136-4139 (1998).

\bibitem{Xu:PRL:2004} P.
Xu, S. H. Ji, S. N. Zhu, X.Q. Yu, J. Sun, H.T.Wang, J. L. He, Y.Y.
Zhu, and N. B. Ming, Phys. Rev. Lett. {\bf 93}, 133904 (2004).



\bibitem{Kurz:JSTQE:02}J.R. Kurz, A.M. Schober, D.S. Hum, A.J. Saltzman, and M.M. Fejer,
IEEE Journal of Selected Topics in Quantum Electron. {\bf 8}, 660-664 (2002).

\bibitem{Ady:OE:2006}  D. Kasimov {\em et al.}, Opt. Express {\bf 14}, 9371 (2006).

\bibitem{Broderick:PRL:2000} N.G.R. Broderick {\em et al.}, Phys. Rev. Lett. {\bf 84}, 4345
(2000).

\bibitem{SaltielOE}S. Saltiel, W. Krolikowski, D.  Neshev, Y.S.  Kivshar, Opt. Express {\bf 15},  4132 (2007).

\bibitem{Freund:PRL:1968}  I. Freund, Phys. Rev. Lett. {\bf 21},
1404 (1968).

\bibitem{FischerAPL2007}
R. Fischer {\em et al.}, Appl. Phys. Lett. {\bf 91}, 031104 (2007).

\bibitem{Ganany}
A. Ganany {\em et al.}, Appl. Phys. B {\bf 85}, 97 (2006).

\bibitem{Zernike}
F. Zernike and J.E. Midwinter, {\em Applied Nonlinear Optics}
(Wiley, New York, 1973).

\bibitem{Landolt}
F. Charra and G.G. Gurzadyan, In: {\em Nonlinear Dielectric
Susceptibilities}, ed. D.F. Nelson (Springer, Berlin, 2000).

\bibitem{bessel}J. Durnin, J. Opt. Soc. Am. A {\bf 4}, 651 (1987): Z. Bouchal and M. Olivik, J.
Mod. Optics {\bf 42}, 1555-1556 (1995).

\bibitem{Nakamura} M. Nakamura {\em et al.}, Jpn. J. Appl. Phys. {\bf 41}, 465 (2002).

\bibitem{Lobov}
A.I. Lobov {\em et al.} http://eprints.soton.ac.uk/42411/

\end{thebibliography}
\end{document}